\documentclass[%
 reprint,
amsmath,
amssymb,
aps, 
pra,
]{revtex4-2}

\usepackage{graphicx} 
\usepackage{dcolumn} 
\usepackage{bm} 
\usepackage{hyperref} 

\usepackage{xcolor}
\usepackage[normalem]{ulem}
\usepackage{amssymb,amsmath,amsfonts}
\usepackage{booktabs}
\usepackage{multirow}
\usepackage{url}

\hypersetup{
    colorlinks = true,
    citecolor = blue,
    linkcolor = blue,
    filecolor = magenta,      
    urlcolor = .,
    }

\begin{document}

\preprint{APS/123-QED}

\title{Time rescaling for second-order feedback-based quantum optimization}

\author{Leticia Bertuzzi}
    \thanks{These authors contributed equally to this work.}
    \affiliation{Departamento de Física, Universidade Federal de Santa Catarina, 88040-900, Florianópolis, Brazil}

\author{João P. Engster}
    \thanks{These authors contributed equally to this work.}
    \affiliation{Departamento de Física, Universidade Federal de Santa Catarina, 88040-900, Florianópolis, Brazil}

\author{Evandro C. R. da Rosa}
    \affiliation{Departamento de Informática e Estatística, Universidade Federal de Santa Catarina, 88040-900, Florianópolis, Brazil}    

\author{Eduardo I. Duzzioni}
    \affiliation{Departamento de Física, Universidade Federal de Santa Catarina, 88040-900, Florianópolis, Brazil}

\date{\today}

\begin{abstract}
Feedback-based quantum algorithms recently gained attention by demonstrating how quantum computers can be used to solve optimization problems without resorting to hybrid quantum-classical architectures. Recent progress made with Feedback-based Algorithm for Quantum Optimization (FALQON) variants significantly reduced the depth of the required circuits. In this work, we merge the time-rescaled and second-order FALQON variants into a unified framework. The results show an important improvement over past modifications, reducing circuit depth and allowing a more flexible choice of time-steps, while maintaining stable solutions. This FALQON variant makes it suitable for the NISQ era.  

\end{abstract}

\maketitle

%
%
\section{Introduction}
\label{sec: introduction}
In recent years we have seen the growth and scaling of quantum processors, with current hardware reaching up to thousands of qubits. Yet, in the so-called NISQ era~\cite{Preskill_2018}, the technological difficulties present in current quantum hardware pose one of the main challenges in the search for quantum advantage~\cite{Huang_2025_perspective, Eisert_2026, Huang_2021_Advantage_learning, Daley_2022_Practical_Advantage, SmithMiles_2025_advantage_challenges}. At this stage, with noisy shallow quantum circuits, hybrid classical-quantum algorithms became the standard approaches. In particular, VQE~\cite{Peruzzo_2014_VQE} and QAOA~\cite{Farhi_2014_QAOA, Zhou_2020_QAOA} stand out as succesful cases. However, hybrid algorithms are subject to the challenges associated with classical optimization and gradient descent methods. More specifically, the existence of barren plateaus is often a roadblock~\cite{Larocca_2025}.

In this scenario, the development of alternatives to hybrid algorithms could provide a means to overcome the issues from classical optimization. In recent years, feedback-based quantum algorithms (FQAs) emerged as candidates by eliminating the need for a classical optimization loop. Inspired by Lyapunov control theory~\cite{Hou_2012}, the FALQON algorithm uses a feedback loop containing only one parameter, estimated from measurements at each layer~\cite{Magann_2022_PRL}. Even though FQAs first appeared in the context of combinatorial optimization, recent applications range from state preparation~\cite{Larsen_2024, Pexe_2024}, excited states calculations~\cite{Rahman_2026}, prime number factorization~\cite{Krishnan_2026}, and DNA assembly problems~\cite{Prado_2026_DNA}.

In spite of being free from classical optimization, FALQON still demands deep circuits, reaching up to thousands of layers for some problems~\cite{Magann_2022_PRL, Pexe_2024}, undermining its use in current quantum hardware. To address this issue, we propose merging two variants of FALQON, each of which showed significant reduction in circuit depth. Namely, the time-rescaling method (TR-FALQON)~\cite{Rattighieri_2025} and the second-order expansion (SO-FALQON)~\cite{Arai_2025}. As a result, we obtain significantly shorter quantum circuits than the original FALQON algorithm~\cite{Magann_2022_PRL}, achieving circuit depths comparable to QAOA while offering superior stability.   

This manuscript is organized as follows: in Sec.~\ref{sec: falqon review} we review the main mathematical aspects of FALQON; in Sec.~\ref{sec: optimizing falqon} we discuss improved variants of the algorithm, focusing on TR-FALQON and SO-FALQON, and combine them into a new, improved version of FALQON. Next, in Sec.~\ref{sec: results} we discuss our results, analysing the MaxCut problem on 3-regular graphs (as in standard FALQON literature) and also on random Erd\H{o}s--R\'enyi graphs. Lastly, we outline conclusions and perspectives in Sec.~\ref{sec: conclusions}.
%
%
\section{FALQON: a brief review}
\label{sec: falqon review}
The algorithm is constructed with respect to the total Hamiltonian~\cite{Magann_2022_PRL}
\begin{equation}
    H = H_p + \beta(t) H_d,
    \label{eq: total hamiltonian}
\end{equation}
in which $H_p$ and $H_d$ are called problem and driver Hamiltonians, respectively. $H_p$ is diagonal in the $Z$-basis, and encodes the solution to the optimization problem of interest in its ground state~\cite{Lucas_2014}. $H_d$ drives the state of the system to explore the solution space, and its effect is modulated by the time-dependent control parameter, $\beta(t)$. 
The algorithm smoothly evolves the quantum system from the ground state of $H_d$ to the ground state of $H_p$, resembling protocols from adiabatic quantum computing~\cite{Farhi_2000, Farhi_2001}. Therefore, at the end of the evolution we require $\beta \rightarrow 0$, rendering $H = H_p$. To achieve this, FALQON builds upon Quantum Lyapunov Control (QLC)~\cite{Isidori_1995, Hou_2012}, defining an appropriate control function $\beta(t)$, which guides the system towards the desired solution. Since the solution is encoded in the ground state of $H_p$, we identify the cost function of the problem as
\begin{equation}
    C(t) \equiv \langle H_p \rangle = \langle \psi(t) | H_p | \psi(t) \rangle.
    \label{eq: cost function}
\end{equation}
To minimize the cost function, we impose that it decreases with time, leading to 
\begin{align}
    \frac{\mathrm{d}}{\mathrm{d}t} \langle  H_p  \rangle = A(t) \beta(t) \leq 0, \quad \forall t,
    \label{eq: cost evolution}
\end{align}
with $A(t) \equiv \langle \psi(t) | i[H_d, H_p] | \psi(t) \rangle$. From Eq.~\eqref{eq: cost evolution} it is clear that an appropriate choice of the control parameter $\beta(t)$ guarantees the monotonic decrease of the cost function. FALQON assures this by choosing the control as~\cite{Magann_2022_PRA}
  \begin{equation}
       \beta(t) = -A(t).
       \label{eq: beta definition}
  \end{equation}

The evolution generated by the total Hamiltonian in Eq.~\eqref{eq: total hamiltonian} is implemented via the Trotter-Suzuki decomposition of the time evolution operator~\cite{Trotter_1959, suzuki_generalized_1976}:
\begin{equation}
    U = U_d (\beta_\ell)U_p \ldots U_d(\beta_1)U_p,
\end{equation}
in which the operators $U_p$ and $U_d$ are given by
\begin{equation}
    \label{eq: evolution operators}
    \begin{aligned}
        U_p &= \exp(-iH_p\Delta t), \\
         U_d(\beta_k) &= \exp(-i\beta_k H_d\Delta t),
    \end{aligned}
\end{equation}
with $\beta_k = \beta[(2k-1)\Delta t]$ for $k = 1,2,...,\ell$, such that $\ell$ is the number of layers and $k$ determines the time step.

To ensure the validity of Eq. \eqref{eq: cost evolution}, $\beta(t)$ is updated as $\beta_{k} =- A_{k-1}$, highlighting the feedback structure of the algorithm: the control parameter is updated with respect to expected values calculated in the previous layer. In particular, by choosing a control parameter based on expected values from previous layers, the algorithm is able to circumvent any classical optimization steps. As usual in trotterized evolutions, the parameter $\Delta t$ must be chosen sufficiently small in order to guarantee proper convergence of the algorithm.
%
%
\section{Optimizing FALQON}
\label{sec: optimizing falqon}
In light of the large circuit depth present in the original formulation of FALQON, several improved versions of the algorithm have been proposed, aiming to reduce circuit depth and improve the quality of the final solutions. Prominent examples include time-rescaled dynamics~\cite{Rattighieri_2025}, a second-order expansion in the energy difference~\cite{Arai_2025}, imaginary-time evolution~\cite{VanLong_2025}, counter-diabatic driving~\cite{Malla2024}, updating previous layers~\cite{Brady_2025}, and measurement-guided strategies~\cite{Rattighieri_2026}. In particular, the latter demonstrates that shallow FALQON circuits can be used to achieve convergence, at the cost of iterating shallow circuit executions (and thus state preparation) several times. In this manuscript, weaim to achieve convergence with few layers, without iteratinge through circuit executions. We do so by merging TR-FALQON~\cite{Rattighieri_2025} and SO-FALQON~\cite{Arai_2025}. Both methods are detailed in the following subsections.
%
%
\subsection{Time-rescaling method}
\label{subsec: time rescaling}
Time-rescaling the evolution of a quantum system has been introduced in Ref.~\cite{Bernardo_2020} and further developed in Ref.~\cite{Ferreira_2024} as a means to achieve shortcuts to adiabaticity. The key idea is to think of time as a parametrized function. This leads to modified dynamics during evolution, while preserving the initial and final states, and enables the target state to be reached faster than usual evolution, in a similar fashion as described in local adiabatic evolution~\cite{Roland_2002}.

Considering a time-dependent Hamiltonian with non-commuting terms, the evolution operator on the interval $[0, t_f]$ is given by
\begin{equation}
    U(t_f,0) = \mathcal{T} \exp \left\{ -\frac{i}{\hbar} \int_0^{t_f} H(t) \, \mathrm{d}t \right\},
\end{equation}
where $\mathcal{T}$ is the time ordering operator. Letting $t = f(\tau)$, we change variables to write the evolution operator as
\begin{equation}
    \mathcal{U}(t_f,0) = \mathcal{T} \exp \left\{ -\frac{i}{\hbar} \int_{f^{-1}(0)}^{f^{-1}(t_f)} H(f(\tau)) \dot{f} (\tau) \, \mathrm{d} \tau \right\},
\end{equation}
which can be interpreted as the evolution operator generated by the modified Hamiltonian
\begin{equation}
    \mathcal{H}(\tau) = H(f(\tau)) \dot{f} (\tau).
\end{equation}
Thus, by appropriately choosing the rescaling function $f(\tau)$, the evolution generated by $\mathcal{H}(\tau)$ can produce the same final state as if evolved according to $H(t)$, however in a shorter time interval. To achieve this goal, some properties should be satisfied~\cite{Bernardo_2020}. The initial times must coincide, i.e., $f^{-1}(0) = 0$; in order to preserve the initial and final states, the original and modified Hamiltonians must be equal at the initial and final times. The latter condition leads to $\dot{f}(0) = \dot{f}(\tau_f) = 1$, with $\tau_f=f^{-1}(t_f)$. Lastly, to shorten the total evolution time, it is required that $\tau_f < t_f$. With those conditions, several rescaling functions can be found, and the usual choices are~\cite{Bernardo_2020, Rattighieri_2025}  
\begin{equation}
    \begin{aligned}
        f_1(\tau) &= a\tau - \frac{t_f}{2\pi a}(a-1) \sin \left( \frac{2\pi a}{t_f} \tau\right), \\
        f_2(\tau) &= \frac{2(a^2-a^3)}{t_f^2} \tau^3 + \frac{3(a^2 - a)}{t_f} \tau^2 + \tau,
        \label{eq: TR functions}
    \end{aligned}
\end{equation}
where $a$ is a constant controlling the magnitude of the rescaling process. Even though such functions are not analytically invertible, from numerical inversion one can check the validity of the aforementioned properties. Moreover, the functions in Eq.~\eqref{eq: TR functions} assure that $\tau_f = t_f/a$. Therefore, if the rescaling factor $a$ is greater (less) than unity, the rescaled protocol produces a faster (slower) evolution.

Given the rescaling protocol described, the trotterized evolution operators in FALQON become
\begin{equation}
    \begin{aligned}
        U_p &= \exp \left( -iH_p\dot{f}_{k}\Delta \tau \right), \\
        U_d(\beta_k) &= \exp \left(-iH_d\beta_{k}\dot{f}_{k}\Delta \tau \right).
    \end{aligned}
    \label{eq: TR operators}
\end{equation}
Thus, the fixed time step $\Delta t$ is replaced by an effective step $\dot{f}_k\Delta \tau$. To determine the new feedback law, we proceed in the same way as the original FALQON, calculating the change in the cost function
\begin{equation}
    \frac{\mathrm{d}}{\mathrm{d} \tau} \langle H_p \rangle = \beta(\tau) A(\tau) \dot{f}(\tau),
\end{equation}
where the quantity $A(\tau)$ is defined analogously. To ensure monotonic decrease in the cost function, the control parameter is redefined as~\cite{Rattighieri_2025}
\begin{equation}
    \beta_k^{\text{(FO-TR)}} = -\frac{A_{k-1}}{\dot{f}_k(\tau)},
    \label{eq: FO-TR beta}
\end{equation}
where FO-TR stands for first order, time rescaled. This choice of the control parameter allows for a monotonic decrease in the cost function, considering a first-order expansion of the evolution operators $U_p$ and $U_d$ in Eq.~\eqref{eq: TR operators}. In the next subsection, we consider the case for the second-order expansion of the evolution operators.
%
%
\subsection{Second-order expansion}
\label{subsec: second-order}
The second-order expansion modifies the feedback law of FALQON, considering a second-order correction in the energy variation between layers $k-1$ and $k$. As a result, a more accurate approximation of the algorithm dynamics is obtained compared to the first-order formulation, allowing for larger time steps and, consequently, fewer layers for the algorithm to converge. In particular, those critical values of $\Delta t$ lead to a reduction in circuit depth of more than one order of magnitude~\cite{Arai_2025}.

The variation of the expected value of the problem Hamiltonian between two consecutive layers can be written as:
\begin{equation}
    \Delta \langle H_p \rangle_k = \langle H_p \rangle_k - \langle H_p \rangle_{k-1}.
\end{equation}
Expanding the difference of the cost function, up to second-order in the time step $\Delta t$, we get
\begin{equation}
    \begin{aligned}
        \Delta \langle H_p \rangle_k &\approx \Delta t \, \beta_k \langle \psi_{k-1} | i[H_d,H_p] |\psi_{k-1} \rangle \\
        & + (\Delta t)^2 \beta_k^2 \langle \psi_{k-1} | \frac{1}{2}[[H_d,H_p],H_d] |\psi_{k-1} \rangle \\
        & + (\Delta t)^2 \beta_k \langle \psi_{k-1} | [[H_d,H_p],H_p] |\psi_{k-1} \rangle \\
        & + \mathcal{O} \left((\Delta t)^3\right).
    \end{aligned}
    \label{eq: SO energy difference}
\end{equation}
Identifying the quantities
\begin{equation}
    \begin{aligned}
        A_{k-1} &= \langle \psi_{k-1}| i[H_d,H_p] |\psi_{k-1}\rangle, \\
        B_{k-1} &= \frac{1}{2} \langle \psi_{k-1}| [[H_d,H_p],H_d] |\psi_{k-1}\rangle, \\
        C_{k-1} &= \langle \psi_{k-1}|[[H_d,H_p],H_p]| \psi_{k-1} \rangle,
    \end{aligned}
    \label{eq: SO expvals}
\end{equation}
and by maximizing $\Delta \langle H_p \rangle_k$ in Eq.~\eqref{eq: SO energy difference} we arrive at the new feedback law
\begin{equation}
    \beta_k^\text{(SO)} = -\frac{A_{k-1} + \Delta t C_{k-1}}{2\Delta t B_{k-1}},
    \label{eq: SO beta}
\end{equation}
where $A_{k-1}$ is the same expectation value used in the first-order cases, $\beta_k^\text{(FO)}$ and $\beta_k^\text{(FO-TR)}$.

It is worth noting that in Eq.~\eqref{eq: SO beta}, $B_{k-1}>0$ must hold. This constraint ensures that the denominator of $\beta_k^\text{(SO)}$ remains well defined and $\langle H_p \rangle_{k} \leq 0 \, \forall \, k$. In Ref.~\cite{Arai_2025}, the authors introduce a cap on $\beta_k^\text{(SO)}$ to avoid possible issues if $B_{k-1}\leq 0$ or if it becomes too small. Whenever that is the case, the first-order feedback law is applied, characterizing a hybrid implementation of the algorithm, alternating between $\beta_k^\text{(SO)}$ and $\beta_k^\text{(FO)}$. Throughout this manuscript, we refer to SO-FALQON considering the hybrid implementation as it renders better results~\cite{Arai_2025}.
%
%
\subsection{Time-rescaled second-order FALQON}
\label{subsec: TR-SO FALQON}
\begin{figure*}
    \centering
    \includegraphics[width=0.99\linewidth]{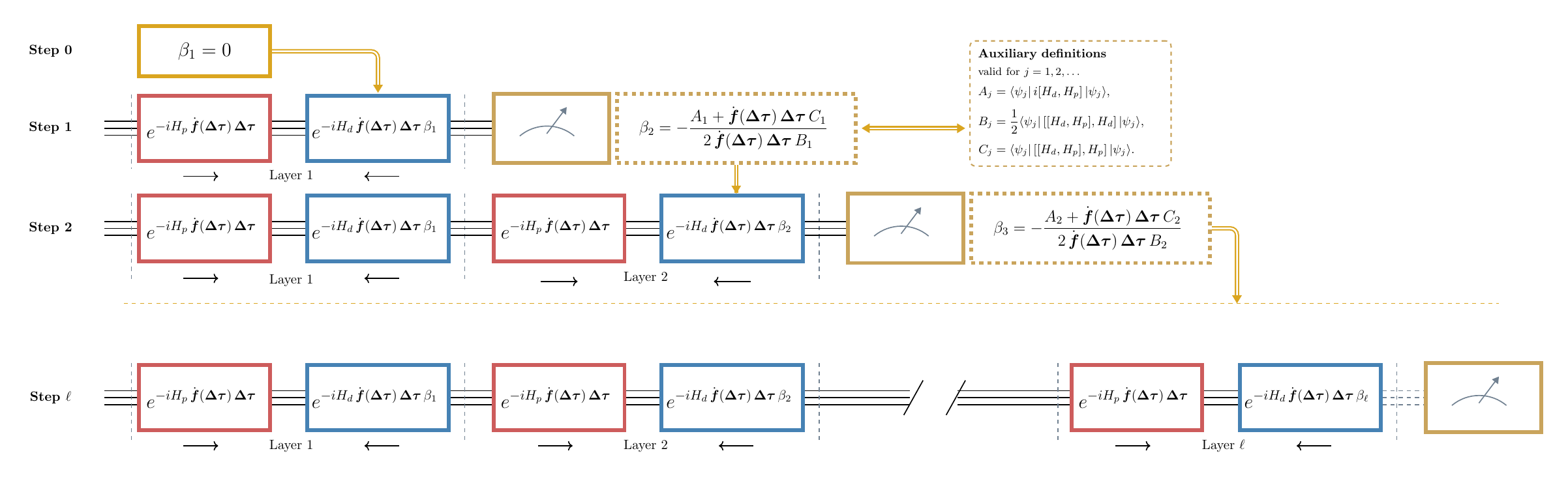} 
    \caption{Circuit model for TR-Hy-FALQON. At the end of each layer, the same quantities required for SO-FALQON are measured and used to calculate the control parameter. The effective time step $\dot{f} \Delta \tau$ is determined according to Eq.~\eqref{eq: TR functions}.}
    \label{fig: circuit model falqon}
\end{figure*}
To combine both approaches, we follow the same steps as Ref.~\cite{Arai_2025}, but considering the time-rescaled scenario from Ref.~\cite{Rattighieri_2025}. In this context, the evolution operators are the same as in Eq.~\eqref{eq: TR operators}. Calculating the energy difference from subsequent layers of TR-FALQON, up to second-order in $\Delta\tau$, we get
\begin{equation}
    \begin{aligned}
        \Delta \langle H_p \rangle_k &\approx \beta_k \dot{f}_k \Delta\tau A_{k-1} \\
        &+ \beta_k^{\,2}\dot{f}_k^{\,2}\Delta\tau^2 B_{k-1} \\
        &+ \beta_k(\dot{f}_k\Delta\tau)^2 C_{k-1},
    \end{aligned}
\end{equation}
where the quantities $A_{k-1}$, $B_{k-1}$, and $C_{k-1}$ are defined in Eq.~\eqref{eq: SO expvals}. Then, the second-order time-rescaled control parameter can be written as
\begin{equation}
    \beta_k^{\text{(SO-TR)}} = - \frac{A_{k-1} + \dot{f}_k\Delta\tau C_{k-1}}{2\dot{f}_k\Delta\tau B_{k-1}}.
\end{equation}
Therefore, the obtained feedback law is analogous to the usual second-order case (see Eq.~\eqref{eq: SO beta}), but with the substitution $\Delta t \rightarrow \dot{f}_k\Delta\tau$. In this sense, the time-rescaling protocol implements an effective time-step of $\dot{f}_k\Delta\tau$.

To summarize, merging SO-FALQON and TR-FALQON produces an algorithm that chooses the control parameter analogously to SO-FALQON, but with the effective time step from TR-FALQON (given by Eq.~\eqref{eq: TR functions}). 

We also keep the same hybrid approach introduced in SO-FALQON: the algorithm selects the smallest value between $\beta_k^\text{(SO-TR)}$ (Eq.~\eqref{eq: SO beta}) and $\beta_k^\text{(FO-TR)}$(Eq.~\eqref{eq: FO-TR beta}). Thus, we denote this version of FALQON by TR-Hy-FALQON. The circuit model for TR-Hy-FALQON is illustrated in Fig.~\ref{fig: circuit model falqon}.
%
\section{Results}
\label{sec: results}
We benchmarked our modifications to FALQON by solving the MaxCut problem on 3-regular graphs and on Erd\H{o}s--R\'enyi graphs, comparing our approach with the standard QAOA and the other FALQON variants discussed in this manuscript. Analyzing 3-regular graphs is typical in the literature of FQAs, while Erd\H{o}s--R\'enyi graphs present less symmetric scenarios, which could lead to more challenging MaxCut instances. Our results were obtained for 3-regular graphs with 12 vertices and Erd\H{o}s--R\'enyi graphs with 14 vertices and parameter $p=0.3$. For all time-rescaled versions of FALQON, we used the first time-rescaling function, $f_1(\tau)$ in Eq.~\eqref{eq: TR functions}, setting $a = 2$. As a figure of merit we adopt the approximation ratio $r$, defined as the quotient between the estimated energy and the ground state energy of $H_p$: 
\begin{equation}
    r = \frac{\langle H_p \rangle}{\langle H_p\rangle_\text{min}}.    
\end{equation} 
All results hereafter were obtained through numerical simulation of the quantum evolution using the Ket quantum programming platform~\cite{Rosa2022Ket, ket2026}.

\begin{figure}
    \centering
    \includegraphics[width=\linewidth]{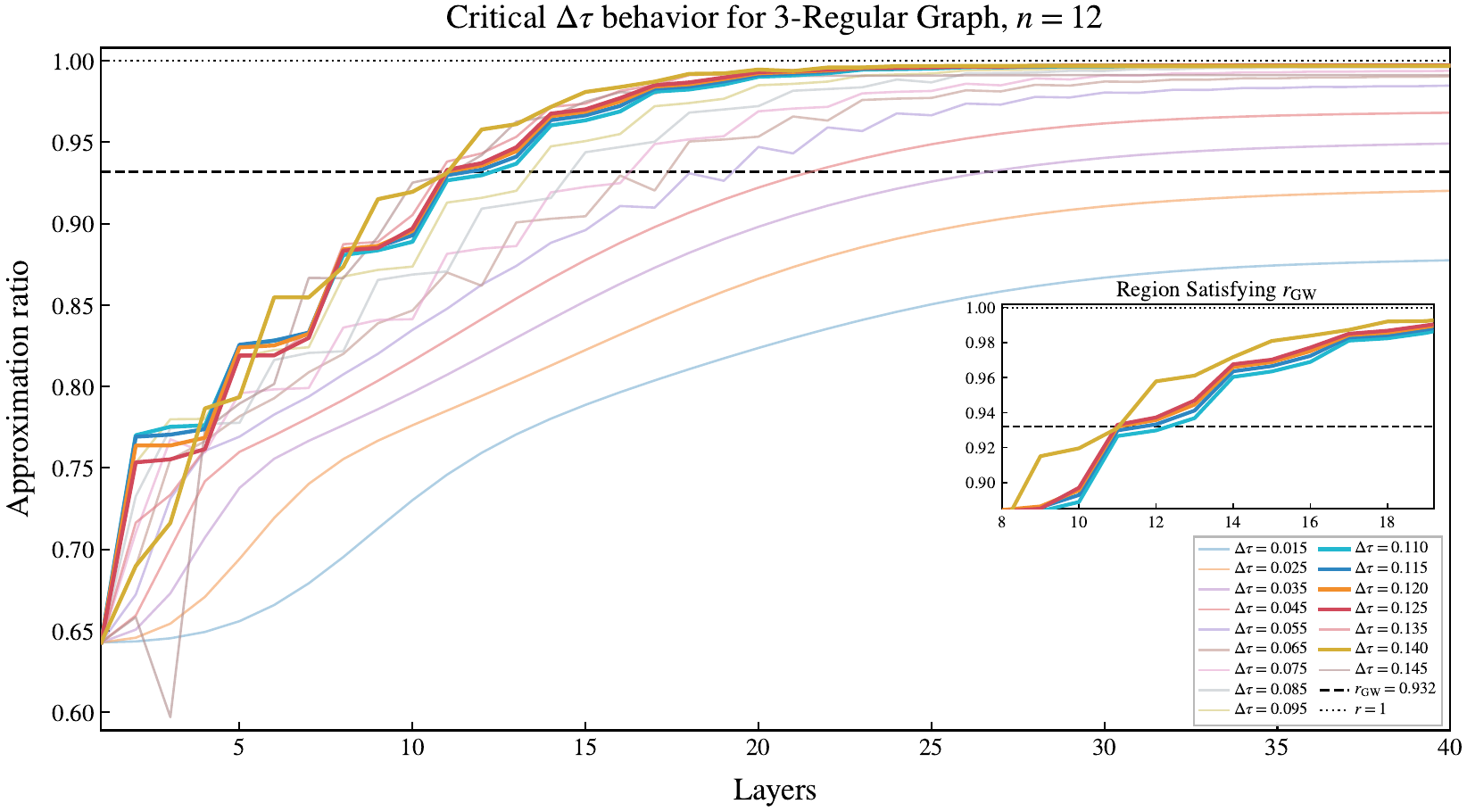} \\
    \includegraphics[width=\linewidth]{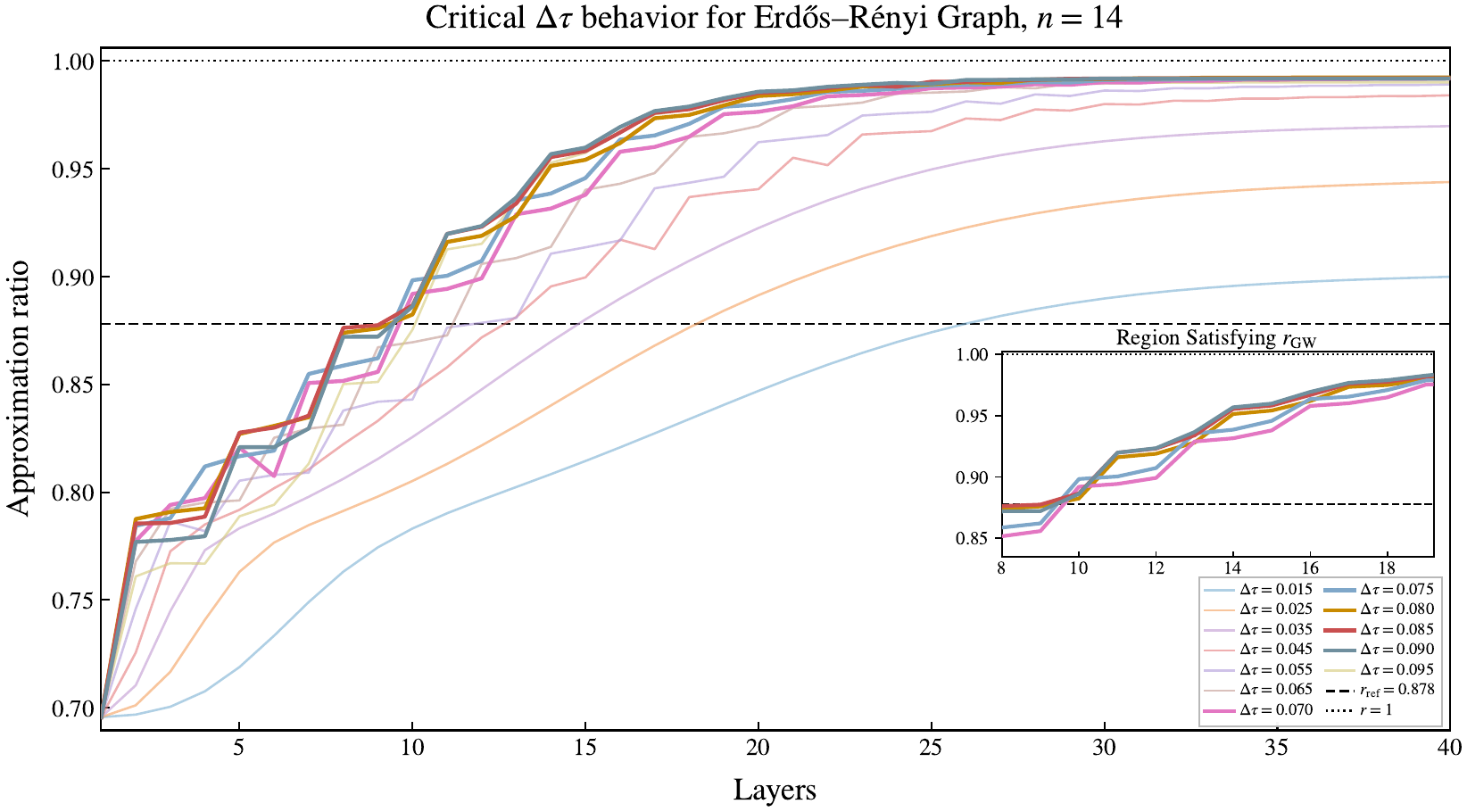}
    \caption{TR-Hy-FALQON approximation ratios for different values of $\Delta \tau$ up to its critical value. The results were obtained for the selected geometries in Fig.~\ref{fig: selected_graphs}, and the bold curves indicate the top five best-performing values for $\Delta \tau$, highlighted in the insets. The horizontal dashed line is the theoretical approximation ratio guarantee provided by the classical Goemans--Williamson algorithm. }
    \label{fig: delta_t comparison}
\end{figure}

\begin{figure}
    \centering
    \includegraphics[width=0.35\linewidth]{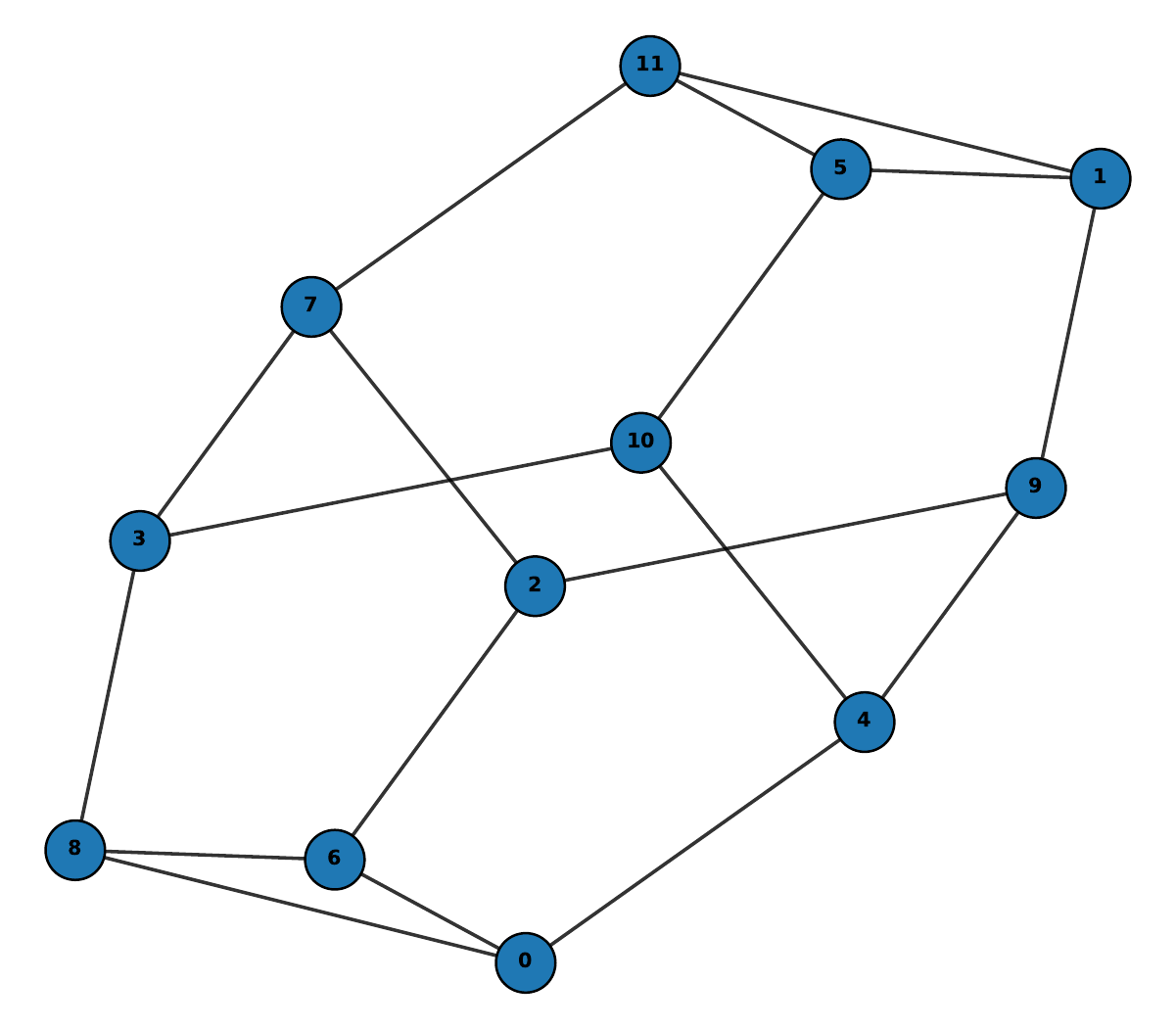}
    \includegraphics[width=0.35\linewidth]{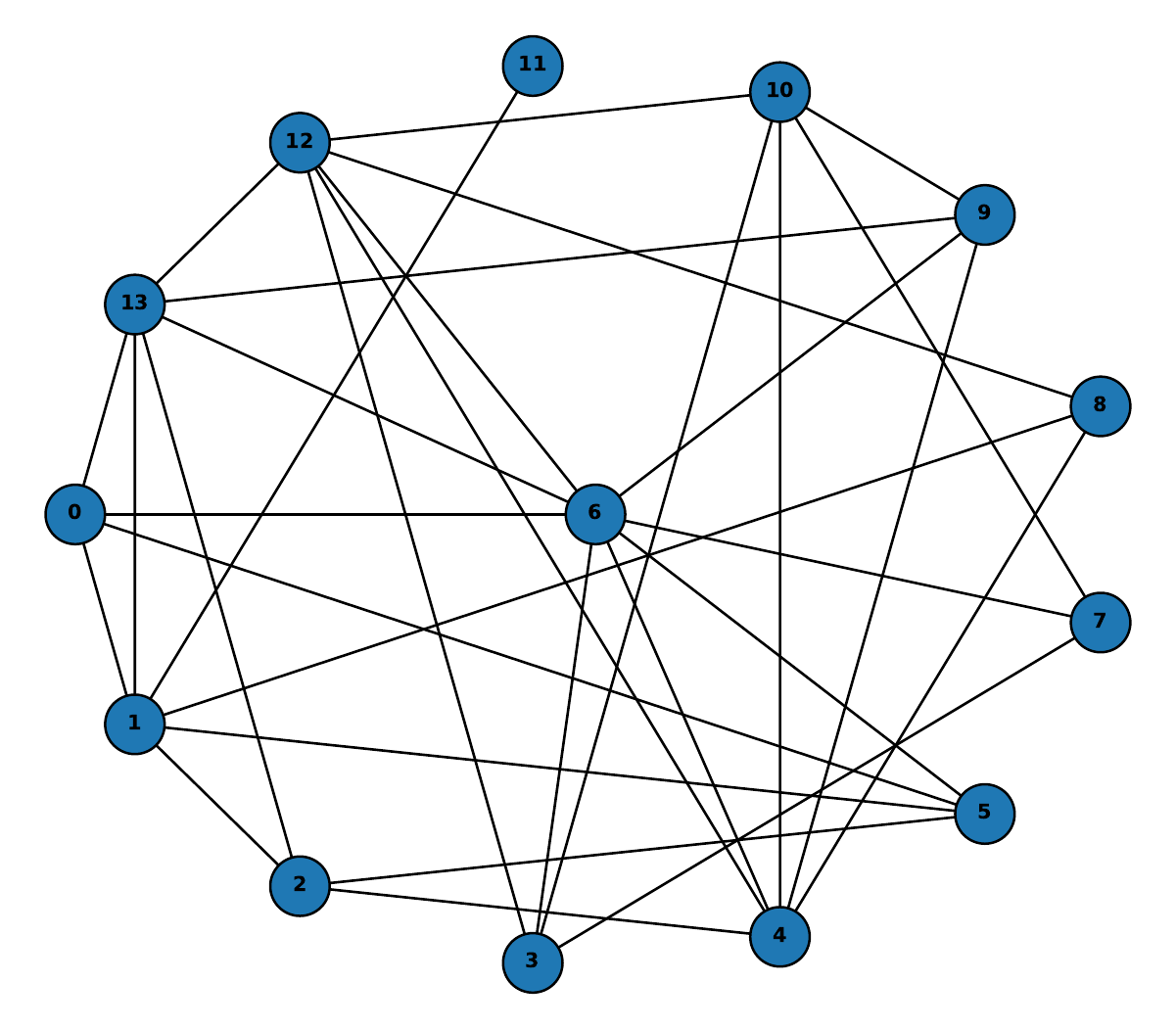}
    \caption{Selected geometries employed to obtain the best approximation ratios for TR-Hy-FALQON described in Fig.~\ref{fig: delta_t comparison}. \textit{Left}: 12-vertex 3-regular graph. \textit{Right}: 14-vertex Erd\H{o}s--R\'enyi graph, with $p=0.3$.}
    \label{fig: selected_graphs}
\end{figure}

In order to properly compare the different FALQON variants, we run the algorithms using the critical values of the parameters $\Delta t$ and $\Delta \tau$, adopted based on values already established in the literature \cite{Magann_2022_PRL, Arai_2025, Rattighieri_2025}. For the TR-Hy-FALQON variant, however, the critical value of $\Delta \tau$ was determined through a numerical sweep over different time-step values, as shown in Fig.~\ref{fig: delta_t comparison}, for the graphs depicted in Fig.~\ref{fig: selected_graphs}. These critical values correspond to the largest admissible time steps that preserve stable convergence while allowing the fastest convergence for each FALQON implementation. It is worth noting that, even for small values of $\Delta \tau$, TR-Hy-FALQON can exhibit non-monotonic behavior -- see the curve with $\Delta \tau = 0.055$ in the top panel of Fig.~\ref{fig: delta_t comparison}. In light of that, we adopt the weaker definition of a critical time step discussed above. This zigzag pattern that appears during part of the evolution time is often related to large time steps, causing the algorithm to oscillate and diverge. Yet, for the second-order variants of FALQON, we observe convergence even if not exclusively monotonic. Furthermore, even though temporal rescaling introduces a multiplicative factor $\dot{f}$ that could be significantly larger than $\Delta \tau$, we find critical values of $\Delta \tau$ close to the ones from SO-FALQON, reported in Ref.~\cite{Arai_2025} for 3-regular graphs.

To evaluate the critical value of $\Delta\tau$ for TR-Hy-FALQON, we adopted $r_{\mathrm{GW}}$ as a reference, based on the classical Goemans--Williamson (GW) algorithm \cite{Goemans-Williamson_1995} for the Max-Cut problem, which provides a theoretical guarantee for the approximation ratio. We adopt $r_{\mathrm{GW}}~ =~0.878$~\cite{Tate2023} as the reference value for the Erd\H{o}s--R\'enyi graphs, whereas $r_{\mathrm{GW}}=0.932$~\cite{Magann_2022_PRL, Arai_2025} was used for the 3-regular graphs.

The values of $\Delta\tau$ presented in the insets of Fig.~\ref{fig: delta_t comparison} correspond to the five best results for each graph geometry, selected by jointly considering the minimum number of layers required to reach the reference value $r_{\mathrm{GW}}$ and the stability of the algorithm throughout the evolution. For the 3-regular graph with $n=12$, these values yielded very similar approximation ratios, all exceeding $0.99$ from the 19th layer onward. In particular, $\Delta\tau=0.140$ reached this threshold earlier, at the 17th layer, and exhibited the best performance in terms of the number of layers required to reach the reference threshold, attaining $r_{\mathrm{GW}}=0.932$ as early as the 11th layer, while the remaining values required a larger number of layers to reach these thresholds. Moreover, this value of $\Delta\tau$ maintained an approximation ratio above $0.996$ after the 23rd layer, without exhibiting divergence or a significant loss in solution quality, and was therefore adopted as $\Delta\tau_c$ for this geometry.

For the Erd\H{o}s--R\'enyi graph with $n=14$, the selected values also yielded similar final approximation ratios. Since all five values reached the reference threshold, $r_{\mathrm{GW}}=0.878$, within 10 layers, with no difference in the number of layers required to reach it, the highest final approximation ratio was used as the main selection criterion. In this case, $\Delta\tau=0.080$ yielded the highest final value, with $r=0.992$, after 32 layers, and was therefore adopted as $\Delta\tau_c$ for this geometry.

\begin{figure*}
    \centering
    \begin{minipage}[t]{0.49\linewidth}
        \centering

        \includegraphics[width=\linewidth]{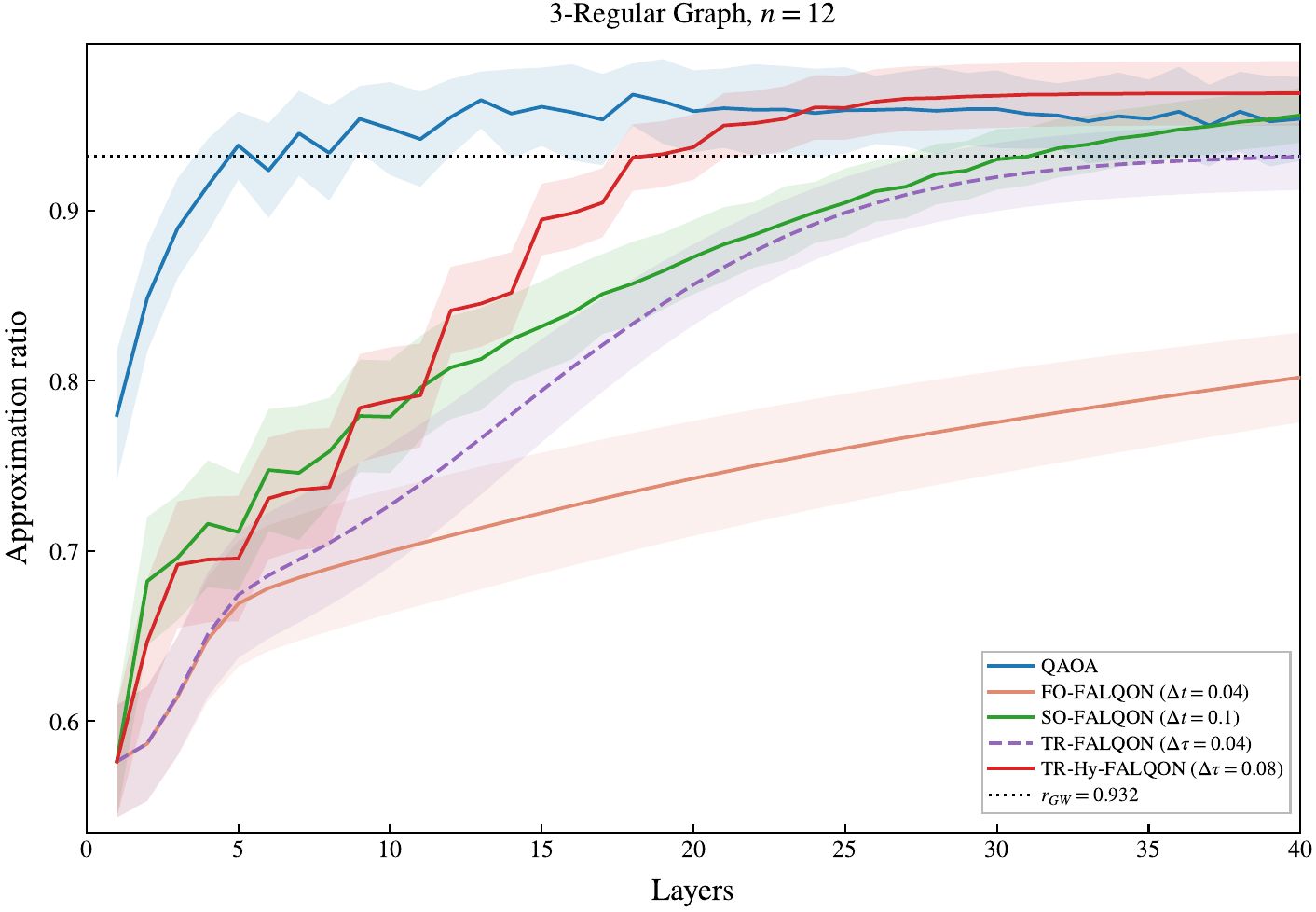}
    \end{minipage}
    \hfill
    \begin{minipage}[t]{0.49\linewidth}
        \centering
        \includegraphics[width=\linewidth]{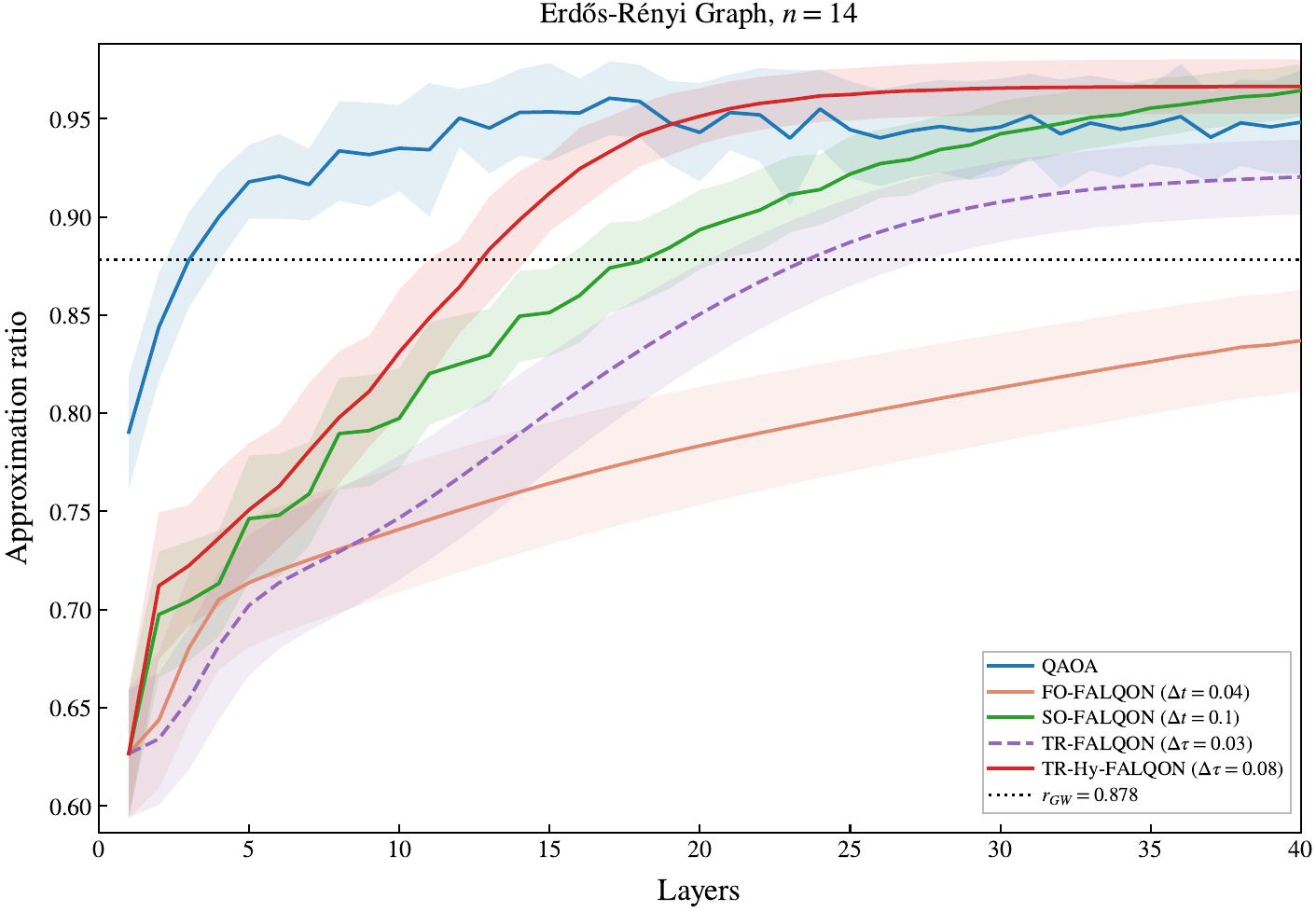}
    \end{minipage}
    \vspace{0.4cm}
    \begin{minipage}{0.49\linewidth}
        \centering
        \includegraphics[width=\linewidth]{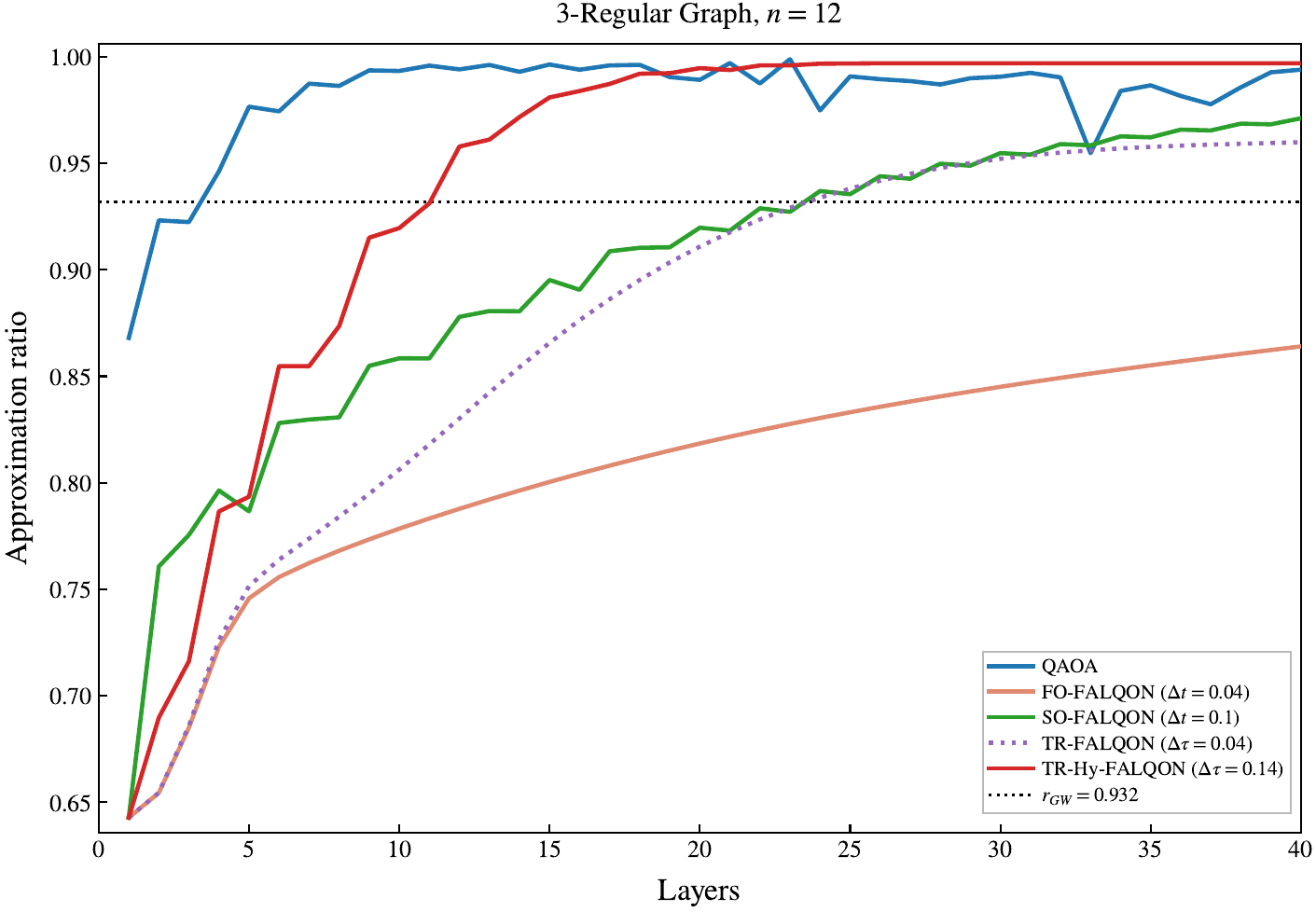}
    \end{minipage}
    \hfill
    \begin{minipage}{0.49\linewidth}
        \centering
        \includegraphics[width=\linewidth]{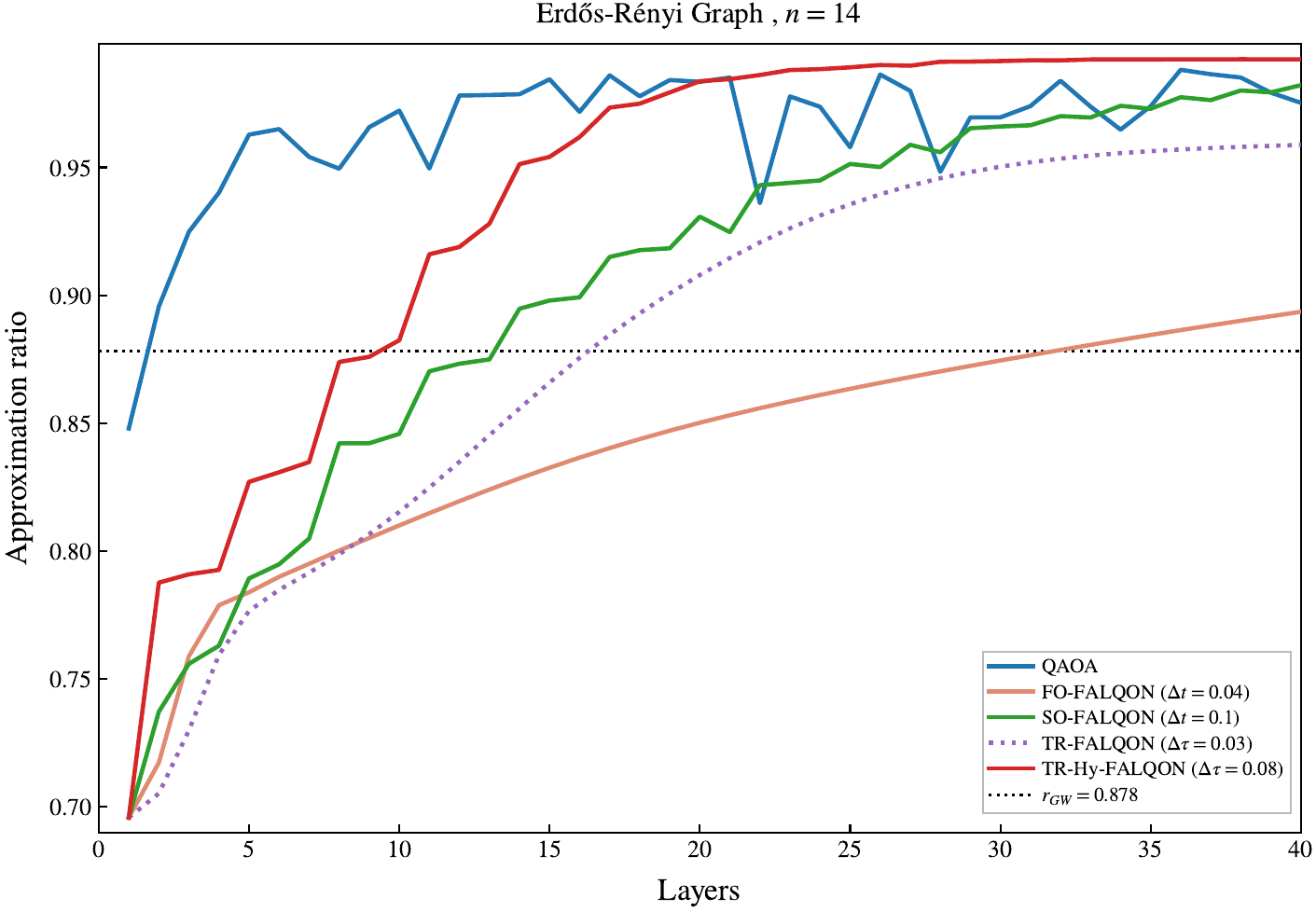}
    \end{minipage}
    \caption{Approximation ratios for random 12-vertex 3-regular graphs and 14-vertex Erd\H{o}s--R\'enyi graphs as a function of the number of layers employed in the quantum circuits generated by the QAOA, FO-FALQON, SO-FALQON, TR-FALQON, and TR-Hy-FALQON algorithms. \textit{Top}: averaged results for 30 different graphs chosen randomly. The shaded regions correspond to the standard deviation. For each FALQON variant, only one critical time-step value, $\Delta t$ or $\Delta \tau$, has been used. \textit{Bottom}: in particular, we used those values of the critical time steps that return the best approximation ratios for the graphs depicted in Fig.~\ref{fig: selected_graphs}. The horizontal dashed line is the theoretical approximation ratio guarantee provided by the classical Goemans--Williamson algorithm. The results indicate how a second-order, time-rescaled version of FALQON accelerates convergence, reaching the same accuracy as QAOA with fewer layers than previous FALQON variants.}
    \label{fig: approx ratio comparison}
\end{figure*}

Fig.~\ref{fig: approx ratio comparison} displays the approximation ratio obtained for the discussed FALQON variants and QAOA. The top panels were obtained by averaging over 30 randomly selected, non-isomorphic, 12-vertex 3-regular graphs (left) and 14-vertex Erd\H{o}s--R\'enyi graphs with $p=0.3$ (right). Compared to the previous FALQON variants, TR-Hy-FALQON (red line) achieved a higher value of the approximation ratio with a reduced number of layers, with similar or better solutions than QAOA at the cost of a few additional layers. Also, these solutions are more stable and present a smaller standard deviation than QAOA. In the bottom panels, we selected the best result among the 30 random regular and Erd\H{o}s--R\'enyi graphs. The topologies of these graphs are depicted in Fig.~\ref{fig: selected_graphs}, and the critical time steps for TR-Hy-FALQON were obtained in Fig.~\ref{fig: delta_t comparison}. The results are summarized in Table~\ref{tab: averaged_performance} (averaged results) and Table~\ref{tab: individual_performance} (selected graphs), where $r_{\mathrm{final}}$ is the approximation ratio obtained at the 40th layer, $\ell^{\star}$ is the number of layers required to reach the $r_{\mathrm{GW}}$. 

\bgroup
\def\arraystretch{1.15}
\begin{table}[ht]
    \centering
    \caption{Performance of the algorithms averaged over 30 graph instances.}
    \label{tab: averaged_performance}
    \begin{ruledtabular}
        \begin{tabular}{llccccc}
        Graph & Metric              &   FO      &   SO      &   TR      &   TR-Hy   &   QAOA \\
        \colrule
        \multirow{3}{*}{3-regular}
        & $r_{\mathrm{final}}$      &   0.802   &   0.955   &   0.931   &   0.968   &   0.953 \\
        & $\ell^{\star}$            &   --      &   32      &   --      &   19      &   5  \\
        & Critical step             &   0.04    &   0.10     &   0.04    &   0.08    &   --  \\
        \midrule
        \multirow{3}{*}{Erd\H{o}s--R\'enyi}
        & $r_{\mathrm{final}}$      &   0.836   &   0.964   &   0.920   &   0.966   &   0.948 \\
        & $\ell^{\star}$            &   --      &   19      &   24      &   13      &   3  \\
        & Critical step             &   0.04    &   0.10    &   0.03    &   0.08    &   --  \\
        \end{tabular}
    \end{ruledtabular}
\end{table}
\egroup
\bgroup
\def\arraystretch{1.15}
\begin{table}[ht]
    \centering
    \caption{Performance of the algorithms for the selected individual graph instances.}
    \label{tab: individual_performance}
    \begin{ruledtabular}
        \begin{tabular}{llccccc}
        Graph & Metric              &   FO      &   SO      &   TR      &   TR-Hy   &   QAOA \\
        \colrule
        \multirow{3}{*}{3-regular}
        & $r_{\mathrm{final}}$      &   0.864   &   0.971   &   0.959   &   0.996   &   0.993 \\
        & $\ell^{\star}$            &   --      &   24      &   24      &   11      &   4  \\
        & Critical step             &   0.04    &   0.10    &   0.04    &   0.14    &   --  \\
        \midrule
        \multirow{3}{*}{Erd\H{o}s--R\'enyi}
        & $r_{\mathrm{final}}$      &   0.893   &   0.982   &   0.958   &   0.992   &   0.975 \\
        & $\ell^{\star}$            &   32      &   14      &   17      &   10      &   2  \\
        & Critical step             &   0.04    &   0.10    &   0.03    &   0.08    &   --  \\
        \end{tabular}
    \end{ruledtabular}
\end{table}
\egroup
In all the cases analyzed, TR-Hy-FALQON shows a significant advantage over the other FALQON variants, reaching $r_{\mathrm{GW}}$ with a considerably smaller number of layers, being the closest to QAOA but displaying solutions with better quality and stability.

Lastly, it is worth mentioning the experimental cost of TR-Hy-FALQON. The quantities needed to evaluate $\beta^\text{(SO)}$ in Eq.~\eqref{eq: SO expvals} introduce the need for extra measurements, which could, in principle, lead to an important overhead. It is clear from Eq.~\eqref{eq: SO expvals} that SO-FALQON requires measuring in the $X$-, $Y$- and $Z$-basis, opposed to the original FALQON (which required $Z$-basis measurements for the cost function and both $Z$- and $Y$-bases for the estimation of $\beta^\text{(FO)}$). However, the additional measurements do not present a setback: as discussed in Ref.~\cite{Arai_2025}, the sampling overhead is not significantly larger than the original FALQON. Since we keep the same feedbak law, the same is valid for TR-Hy-FALQON.
%
%
\section{Conclusion}
\label{sec: conclusions}
We propose a new variant of FALQON, named TR-Hy-FALQON, which merges the time rescaling method and the second order expansion from Ref.~\cite{Rattighieri_2025} and Ref.~\cite{Arai_2025}, respectively. The results demonstrate improvements in the performance of the algorithm for both classes of graphs analyzed, 3-regular and Erd\H{o}s--R\'enyi graphs. Among the FALQON variants, TR-Hy-FALQON achieved the highest approximation ratios and surpassed $r_\mathrm{GW}$ with fewer layers, exhibiting a behavior closer to that of QAOA. Another favorable characteristics of TR-Hy-FALQON is the stability of the solutions, a property not shared by QAOA.    

In addition, the proposed approach provides greater flexibility in the choice of the time step $\Delta\tau$, while maintaining stable behavior even for comparatively larger values of $\Delta\tau$. This allows for larger time increments, accelerating convergence without compromising the quality of the solution. This aspect is particularly relevant for current quantum devices, in which errors and noise can accumulate throughout circuit execution.

By achieving high-quality solutions with shallower circuits, TR-Hy-FALQON points to a promising alternative for quantum optimization in the NISQ era, improving the performance of FQAs. With our modifications, the convergence of FALQON approaches that of QAOA while retaining the main advantages of FQAs: avoiding classical optimization of its parameters and relying only on measured quantities.
%
%
\begin{acknowledgments}

The authors gratefully acknowledge the financial support provided by the Brazilian funding agencies Coordenação de Aperfeiçoamento de Pessoal de Nível Superior (CAPES) and Conselho Nacional de Desenvolvimento Científico e Tecnológico (CNPq) through Grants No. 409673/2022-6 and 408341/2022-0, by the National Institute of Science and Technology for Quantum Information
(INCT-IQ) under Grant No. 465469/2014-0, and by the
National Institute of Science and Technology for Applied Quantum Computing (INCT-CQA) under Process No. 408884/2024-0. The authors also thank Dr. Keisuke Fujii for bringing to their attention, during a conference, the work that motivated and contributed to the development of this research.

\end{acknowledgments}
\bibliography{apssamp}

@PREAMBLE{
 "\providecommand{\noopsort}[1]{}" 
 # "\providecommand{\singleletter}[1]{#1}%" 
}

@article{Larocca_2025,
   title={Barren plateaus in variational quantum computing},
   volume={7},
   ISSN={2522-5820},
   url={http://dx.doi.org/10.1038/s42254-025-00813-9},
   DOI={10.1038/s42254-025-00813-9},
   number={4},
   journal={Nature Reviews Physics},
   publisher={Springer Science and Business Media LLC},
   author={Larocca, Martín and Thanasilp, Supanut and Wang, Samson and Sharma, Kunal and Biamonte, Jacob and Coles, Patrick J. and Cincio, Lukasz and McClean, Jarrod R. and Holmes, Zoë and Cerezo, M.},
   year={2025},
   month=Mar, pages={174–189} }

@misc{Farhi_2014_QAOA,
      title={A Quantum Approximate Optimization Algorithm}, 
      author={Edward Farhi and Jeffrey Goldstone and Sam Gutmann},
      year={2014},
      eprint={1411.4028},
      archivePrefix={arXiv},
      url={https://arxiv.org/abs/1411.4028}, 
}

@misc{Farhi_2000,
      title={Quantum Computation by Adiabatic Evolution}, 
      author={Edward Farhi and Jeffrey Goldstone and Sam Gutmann and Michael Sipser},
      year={2000},
      eprint={quant-ph/0001106},
      archivePrefix={arXiv},
      url={https://arxiv.org/abs/quant-ph/0001106}, 
}

@article{Farhi_2001,
   title={A Quantum Adiabatic Evolution Algorithm Applied to Random Instances of an {NP}-Complete Problem},
   volume={292},
   ISSN={1095-9203},
   url={http://dx.doi.org/10.1126/science.1057726},
   DOI={10.1126/science.1057726},
   number={5516},
   journal={Science},
   publisher={American Association for the Advancement of Science (AAAS)},
   author={Farhi, Edward and Goldstone, Jeffrey and Gutmann, Sam and Lapan, Joshua and Lundgren, Andrew and Preda, Daniel},
   year={2001},
   month=apr, 
   pages={472–475}
}

@article{Hou_2012,
	title = {Optimal {Lyapunov}-based quantum control for quantum systems},
	volume = {86},
	issn = {1050-2947, 1094-1622},
	url = {http://arxiv.org/abs/1203.5373},
	doi = {10.1103/PhysRevA.86.022321},
	number = {2},
	urldate = {2024-09-23},
	journal = {Physical Review A},
	author = {Hou, S. C. and Khan, M. A. and Dong, Daoyi and Petersen, Ian R. and Yi, X. X.},
	month = aug,
	year = {2012},
	note = {arXiv:1203.5373 [quant-ph]},
	pages = {022321},
}

@book{Isidori_1995,
	address = {London},
	series = {Communications and {Control} {Engineering}},
	title = {Nonlinear {Control} {Systems}},
	copyright = {http://www.springer.com/tdm},
	isbn = {9781447139096 9781846286155},
	url = {http://link.springer.com/10.1007/978-1-84628-615-5},
	urldate = {2024-09-24},
	publisher = {Springer},
	author = {Isidori, Alberto},
	editor = {Sontag, E. D. and Thoma, M. and Isidori, A. and Van Schuppen, J. H.},
	year = {1995},
	doi = {10.1007/978-1-84628-615-5},
}

@article{Pexe_2024,
  title = {Using a feedback-based quantum algorithm to analyze the critical properties of the ANNNI model without classical optimization},
  author = {Pexe, G. E. L. and Rattighieri, L. A. M. and Malvezzi, A. L. and Fanchini, F. F.},
  journal = {Phys. Rev. B},
  volume = {110},
  issue = {22},
  pages = {224422},
  numpages = {13},
  year = {2024},
  month = {Dec},
  publisher = {American Physical Society},
  doi = {10.1103/PhysRevB.110.224422},
  url = {https://link.aps.org/doi/10.1103/PhysRevB.110.224422}
}

@article{Larsen_2024,
	title = {Feedback-based quantum algorithms for ground state preparation},
	volume = {6},
	issn = {2643-1564},
	url = {https://link.aps.org/doi/10.1103/PhysRevResearch.6.033336},
	doi = {10.1103/PhysRevResearch.6.033336},
	number = {3},
	urldate = {2025-02-03},
	journal = {Physical Review Research},
	author = {Larsen, James B. and Grace, Matthew D. and Baczewski, Andrew D. and Magann, Alicia B.},
	month = sep,
	year = {2024},
	pages = {033336},
}

@article{Malla2024,
  title = {Feedback-based quantum algorithm inspired by counterdiabatic driving},
  author = {Malla, Rajesh K. and Sukeno, Hiroki and Yu, Hongye and Wei, Tzu-Chieh and Weichselbaum, Andreas and Konik, Robert M.},
  journal = {Phys. Rev. Res.},
  volume = {6},
  issue = {4},
  pages = {043068},
  numpages = {15},
  year = {2024},
  month = {Oct},
  publisher = {American Physical Society},
  doi = {10.1103/PhysRevResearch.6.043068},
  url = {https://link.aps.org/doi/10.1103/PhysRevResearch.6.043068}
}

@article{Rattighieri_2025,
   title={Accelerating feedback-based quantum algorithms through time rescaling},
   volume={112},
   ISSN={2469-9934},
   url={http://dx.doi.org/10.1103/qc91-5mj2},
   DOI={10.1103/qc91-5mj2},
   number={4},
   journal={Physical Review A},
   publisher={American Physical Society (APS)},
   author={Rattighieri, L. A. M. and Pexe, G. E. L. and Bernardo, B. L. and Fanchini, F. F.},
   year={2025},
   month=Oct }

@misc{Rattighieri_2026,
      title={Measurement-Guided State Refinement for Shallow Feedback-Based Quantum Optimization Algorithm}, 
      author={Lucas A. M. Rattighieri and Pedro M. Prado and Marcos C. de Oliveira and Felipe F. Fanchini},
      year={2026},
      eprint={2602.20407},
      archivePrefix={arXiv},
      primaryClass={quant-ph},
      url={https://arxiv.org/abs/2602.20407}, 
}

@article{Magann_2022_PRL,
  title = {Feedback-Based Quantum Optimization},
  author = {Magann, Alicia B. and Rudinger, Kenneth M. and Grace, Matthew D. and Sarovar, Mohan},
  journal = {Phys. Rev. Lett.},
  volume = {129},
  issue = {25},
  pages = {250502},
  numpages = {7},
  year = {2022},
  month = {Dec},
  publisher = {American Physical Society},
  doi = {10.1103/PhysRevLett.129.250502},
}

@article{Magann_2022_PRA,
   title={Lyapunov-control-inspired strategies for quantum combinatorial optimization},
   volume={106},
   ISSN={2469-9934},
   url={http://dx.doi.org/10.1103/PhysRevA.106.062414},
   DOI={10.1103/physreva.106.062414},
   number={6},
   journal={Physical Review A},
   publisher={American Physical Society (APS)},
   author={Magann, Alicia B. and Rudinger, Kenneth M. and Grace, Matthew D. and Sarovar, Mohan},
   year={2022},
   month=Dec }

@article{Arai_2025,
   title={Scalable circuit depth reduction in feedback-based quantum optimization with a quadratic approximation},
   volume={7},
   ISSN={2643-1564},
   url={http://dx.doi.org/10.1103/PhysRevResearch.7.013035},
   DOI={10.1103/physrevresearch.7.013035},
   number={1},
   journal={Physical Review Research},
   publisher={American Physical Society (APS)},
   author={Arai, Don and Okada, Ken N. and Nakano, Yuichiro and Mitarai, Kosuke and Fujii, Keisuke},
   year={2025},
   month=Jan }

@article{Brady_2025,
  title = {Feedback-based optimally controlled quantum states},
  author = {Brady, Lucas T. and Hadfield, Stuart},
  journal = {Phys. Rev. A},
  volume = {111},
  issue = {6},
  pages = {062406},
  numpages = {12},
  year = {2025},
  month = {Jun},
  publisher = {American Physical Society},
  doi = {10.1103/PhysRevA.111.062406},
  url = {https://link.aps.org/doi/10.1103/PhysRevA.111.062406}
}

@misc{VanLong_2025,
      title={Imaginary-time-enhanced feedback-based quantum algorithms for universal ground-state preparation}, 
      author={Thanh Nguyen Van Long and Lan Nguyen Tran and Le Bin Ho},
      year={2025},
      eprint={2512.13044},
      archivePrefix={arXiv},
      primaryClass={quant-ph},
      url={https://arxiv.org/abs/2512.13044}, 
}

@article{Bernardo_2020,
  title = {Time-rescaled quantum dynamics as a shortcut to adiabaticity},
  author = {Bernardo, Bert\'ulio de Lima},
  journal = {Phys. Rev. Res.},
  volume = {2},
  issue = {1},
  pages = {013133},
  numpages = {7},
  year = {2020},
  month = {Feb},
  publisher = {American Physical Society},
  doi = {10.1103/PhysRevResearch.2.013133},
  url = {https://link.aps.org/doi/10.1103/PhysRevResearch.2.013133}
}

@misc{Ferreira_2024,
      title={Shortcuts to adiabaticity designed via time-rescaling follow the same transitionless route}, 
      author={J. L. Montenegro Ferreira and Ângelo F. da Silva França and Alexandre Rosas and Bertúlio de Lima Bernardo},
      year={2024},
      eprint={2406.07433},
      archivePrefix={arXiv},
      primaryClass={quant-ph},
      url={https://arxiv.org/abs/2406.07433}, 
}

@article{Roland_2002,
  title = {Quantum search by local adiabatic evolution},
  author = {Roland, J\'er\'emie and Cerf, Nicolas J.},
  journal = {Phys. Rev. A},
  volume = {65},
  issue = {4},
  pages = {042308},
  numpages = {6},
  year = {2002},
  month = {Mar},
  publisher = {American Physical Society},
  doi = {10.1103/PhysRevA.65.042308},
  url = {https://link.aps.org/doi/10.1103/PhysRevA.65.042308}
}

@article{Preskill_2018,
   title={Quantum Computing in the NISQ era and beyond},
   volume={2},
   ISSN={2521-327X},
   url={http://dx.doi.org/10.22331/q-2018-08-06-79},
   DOI={10.22331/q-2018-08-06-79},
   journal={Quantum},
   publisher={Verein zur Forderung des Open Access Publizierens in den Quantenwissenschaften},
   author={Preskill, John},
   year={2018},
   month=Aug, pages={79} }

@misc{Prado_2026_DNA,
      title={Quantum feedback algorithms for DNA assembly using FALQON variants}, 
      author={Pedro M. Prado and Lucas A. M. Rattighieri and Rafael Simões do Carmo and Giovanni S. Franco and Guilherme E. L. Pexe and Alexandre Drinko and Erick G. Dorlass and Tatiana F. de Almeida and Felipe F. Fanchini},
      year={2026},
      eprint={2602.21080},
      archivePrefix={arXiv},
      primaryClass={quant-ph},
      url={https://arxiv.org/abs/2602.21080}, 
}

@article{Huang_2025_perspective,
  title = {Vast World of Quantum Advantage},
  author = {Huang, Hsin-Yuan and Choi, Soonwon and McClean, Jarrod R. and Preskill, John},
  journal = {Phys. Rev. X},
  volume = {16},
  issue = {3},
  pages = {030501},
  numpages = {29},
  year = {2026},
  month = {Jul},
  publisher = {American Physical Society},
  doi = {10.1103/tn89-g1xz},
  url = {https://link.aps.org/doi/10.1103/tn89-g1xz}
}

@article{Rahman_2026,
  author={Abdul Rahman, Salahuddin and Karabacak, {\"O}zkan and Wisniewski, Rafal},
  journal={IEEE Transactions on Quantum Engineering}, 
  title={Feedback-Based Quantum Algorithm for Excited States Calculation}, 
  year={2026},
  volume={7},
  number={},
  pages={1-16},
  doi={10.1109/TQE.2026.3654528}
  }

@article{Krishnan_2026,
  title = {Experimental prime factorization via feedback quantum control},
  author = {Krishnan, K. B. Hari and Varma, Vishal and Mahesh, T. S.},
  journal = {Phys. Rev. A},
  volume = {114},
  issue = {2},
  pages = {022413},
  numpages = {9},
  year = {2026},
  month = {Aug},
  publisher = {American Physical Society},
  doi = {10.1103/mx49-b924},
  url = {https://link.aps.org/doi/10.1103/mx49-b924}
}

@misc{Eisert_2026,
      title={Mind the gaps: The fraught road to quantum advantage}, 
      author={Jens Eisert and John Preskill},
      year={2026},
      eprint={2510.19928},
      archivePrefix={arXiv},
      primaryClass={quant-ph},
      url={https://arxiv.org/abs/2510.19928}, 
}

@article{Daley_2022_Practical_Advantage,
	title = {Practical quantum advantage in quantum simulation},
	volume = {607},
	issn = {1476-4687},
	url = {https://doi.org/10.1038/s41586-022-04940-6},
	doi = {10.1038/s41586-022-04940-6},
	number = {7920},
	journal = {Nature},
	author = {Daley, Andrew J. and Bloch, Immanuel and Kokail, Christian and Flannigan, Stuart and Pearson, Natalie and Troyer, Matthias and Zoller, Peter},
	month = jul,
	year = {2022},
	pages = {667--676},
}

@article{Huang_2021_Advantage_learning,
  title={Quantum advantage in learning from experiments},
  author={Hsin-Yuan Huang and Mick Broughton and Jordan S. Cotler and Sitan Chen and Jerry Zheng Li and Masoud Mohseni and Hartmut Neven and Ryan Babbush and Richard Kueng and John Preskill and Jarrod R. McClean},
  journal={Science},
  year={2021},
  volume={376},
  pages={1182 - 1186},
  url={https://api.semanticscholar.org/CorpusID:244799643}
}

@article{SmithMiles_2025_advantage_challenges,
doi = {10.1088/2058-9565/add61d},
url = {https://doi.org/10.1088/2058-9565/add61d},
year = {2025},
month = {may},
publisher = {IOP Publishing},
volume = {10},
number = {3},
pages = {033001},
author = {Smith-Miles, Kate A and Hoos, Holger H and Wang, Hao and Bäck, Thomas and Osborne, Tobias J},
title = {The travelling salesperson problem and the challenges of near-term quantum advantage},
journal = {Quantum Science and Technology},
}

@article{Peruzzo_2014_VQE,
   title={A variational eigenvalue solver on a photonic quantum processor},
   volume={5},
   ISSN={2041-1723},
   url={http://dx.doi.org/10.1038/ncomms5213},
   DOI={10.1038/ncomms5213},
   number={1},
   journal={Nature Communications},
   publisher={Springer Science and Business Media LLC},
   author={Peruzzo, Alberto and McClean, Jarrod and Shadbolt, Peter and Yung, Man-Hong and Zhou, Xiao-Qi and Love, Peter J. and Aspuru-Guzik, Alán and O’Brien, Jeremy L.},
   year={2014},
   month=jun}

@article{Zhou_2020_QAOA,
   title={Quantum Approximate Optimization Algorithm: Performance, Mechanism, and Implementation on Near-Term Devices},
   volume={10},
   ISSN={2160-3308},
   url={http://dx.doi.org/10.1103/PhysRevX.10.021067},
   DOI={10.1103/physrevx.10.021067},
   number={2},
   journal={Physical Review X},
   publisher={American Physical Society (APS)},
   author={Zhou, Leo and Wang, Sheng-Tao and Choi, Soonwon and Pichler, Hannes and Lukin, Mikhail D.},
   year={2020},
   month=jun}

@article{Lucas_2014,
   title={Ising formulations of many NP problems},
   volume={2},
   ISSN={2296-424X},
   url={http://dx.doi.org/10.3389/fphy.2014.00005},
   DOI={10.3389/fphy.2014.00005},
   journal={Frontiers in Physics},
   publisher={Frontiers Media SA},
   author={Lucas, Andrew},
   year={2014} }

@article{Goemans-Williamson_1995,
    author = {Goemans, Michel X. and Williamson, David P.},
    title = {Improved approximation algorithms for maximum cut and satisfiability problems using semidefinite programming},
    year = {1995},
    issue_date = {Nov. 1995},
    publisher = {Association for Computing Machinery},
    address = {New York, NY, USA},
    volume = {42},
    number = {6},
    issn = {0004-5411},
    url = {https://doi.org/10.1145/227683.227684},
    doi = {10.1145/227683.227684},
    journal = {J. ACM},
    month = nov,
    pages = {1115–1145},
    numpages = {31},
}

@article{Tate2023,
  author  = {Tate, Reuben and Moondra, Jai and Gard, Bryan and Mohler, Greg and Gupta, Swati},
  title   = {Warm-Started QAOA with Custom Mixers Provably Converges and Computationally Beats Goemans-Williamson's Max-Cut at Low Circuit Depths},
  journal = {Quantum},
  volume  = {7},
  pages   = {1121},
  year    = {2023},
  doi     = {10.22331/q-2023-09-26-1121}
}

@article{Rosa2022Ket,
  author  = {Evandro Chagas Ribeiro Da Rosa and Rafael De Santiago},
  title   = {Ket Quantum Programming},
  journal = {ACM Journal on Emerging Technologies in Computing Systems},
  volume  = {18},
  number  = {1},
  pages   = {1--25},
  year    = {2022},
  doi     = {10.1145/3474224}
}

@article{ket2026,
    title = {Full Quantum Stack: Ket Platform},
    doi = {10.1007/s13538-025-01981-w},
    volume = {56},
    issn = {0103-9733, 1678-4448},
    language = {en},
    number = {1},
    urldate = {2026-01-08},
    journal = {Brazilian Journal of Physics},
    author = {Rosa, Evandro and Lussi, Eduardo and Marchi, Jerusa and De Santiago, Rafael and Duzzioni, Eduardo},
    month = feb,
    year = {2026},
    pages = {45},
}

@article{suzuki_generalized_1976,
    title = {Generalized {Trotter}'s formula and systematic approximants of exponential operators and inner derivations with applications to many-body problems},
    volume = {51},
    issn = {1432-0916},
    url = {https://doi.org/10.1007/BF01609348},
    doi = {10.1007/BF01609348},
    language = {en},
    number = {2},
    urldate = {2026-09-10},
    journal = {Communications in Mathematical Physics},
    author = {Suzuki, Masuo},
    month = jun,
    year = {1976},
    pages = {183--190},
}

@article{Trotter_1959,
   title={On the Product of Semi-Groups of Operators},
   volume={10},
   ISSN={2296-424X},
   url={http://dx.doi.org/10.3389/fphy.2014.00005},
   DOI={10.1090/S0002-9939-1959-0108732-6},
   journal={Proceedings of the American Mathematical Society},
   author={Trotter, H.F.},
   pages={545-551},
   year={1959}
   }
\end{document}